# Towards a facile method to protect shorelines

Daniel Peterson[1]

*University of Washington, Seattle, Washington, 98195, USA*

Robert Segmaier[2]

*University of Washington, Seattle, Washington, 98195, USA*

Sarah Palmer,[3]
*University of Washington, Seattle, Washington, 98195, USA*

## I. Abstract

**Mangroves are found worldwide across the rivers and coastlines in tropical regions. They are robust against storm surges and tsunami for a long time. The roots have the most contributions for their resiliency and therefore can be inspired for future manmade structures. The motion of water in riverine mangrove forest is expected to be impacted by mangrove roots, which in turn disturb the transport of nutrients, contaminants, and residues in these systems. In this paper, a facile method for protecting shoreline is described and review and the significant impact of this method were reviewed. Bioinspired simplified models as and obstruction to water currents in shorelines, and coastal areas are presented. It was found that Mangrove roots produce complex flow structure interactions with their environment, which affect the nutrient, habitat and aquatic animals. Analysis of the flow structure behind the roots extend to a broad range of mangrove-inspired applications and provide understandings into flows that are more complex and more indicative of the flows encountered in the unidirectional riverine flow.**

## II. Introduction

Red mangrove aids control the tidal hydrodynamics of uni-directional riverine flow. Red mangrove roots entail of the robust network that can withstand harsh circumstances and interact with nutrient and sediment [1]–[3]. It generates an irregular flow pattern in the tidal currents in both the tidal creeks and the mangrove swamps. A thoughtful comprehension of the hydrodynamics of mangrove roots facilitates the incorporation of bio-inspired mangrove structure for erosion control, coastal protection, and habitat reconstruction. In the region of less mangrove vegetation, catastrophic coastal destruction can happen, however, zones with shoreline tree vegetation are obviously less damaged than areas without plants[4]. Mangrove swamp is the most noteworthy coastal tree vegetation in the coastal zones and can be used as a facile method to protect against harsh conditions.

Although the Atlantic Ocean has relatively less steep shorelines and a uniform beach profile, the shoreline embraces vegetated as well as no-vegetated areas [3], [5]. At the river banks, the flow of uninhabited parts of a village and removed a sand spit that previously blocked the river. Nonetheless, areas with mangroves were significantly less devastated than other areas. Damages to communities also varied distinctly; in the north of Atlantic ocean, stands of mangroves had five related villages, two on the coast and three behind the mangrove [6], [7]. The communities in the

---

coastal areas were destroyed, but those behind the mangrove hurt no damage even if the currents broken areas unprotected by vegetation north and south of these communities[2], [4].

The mangrove root system is a multifarious combination of roots, trunk, and branches. Its arrangement varies from the bed lowest part to the top canopy. Mangrove roots, Rhizophora species, are the most typical feature of mangroves and are categorized based on their geometrical properties. Wolanski et al. [8] made two-dimensional outlines of these roots. The segment of a cross-sectional area between two trees that were obstructed by prop roots was assessed from the sketches. He found that the congested area decreased promptly with root height. Mazda et al. [9]obtained dimensional quantities of tree trunks, roots and their geometrical features, such as root elevations and root sizes of a number of mangroves in the field. The porosity of the mangrove roots was estimated based on the inundated root volume to the total volume to calculate the drag force made by the presence of the roots. Wolanski et al.[10] measured the mass density of the mangroves and indicated that the wood of the mangrove trunk has a higher density than water, but dry mangrove wood, which was reported by Sharps et al. [11] with a lower density floats in seawater. Sharps et al.. reported the densities of mangrove wood and they showed that mangrove wood is generally dense so resistant to marine weakening. Nevertheless, the mechanical performance of mangrove roots received less consideration, which caused difficulties in simulating the effects of mangrove roots in physical experimental models.

It is essential to characterize hydrodynamics under different flow settings to enumerate the flow pattern interaction with the mangrove models that allow changing factors such as porosity and flexibility. The motivation of this work is to review the recent characterization works for one and two dimensional flow feature past mangrove root models. For this efforts, roots modeling and hydrodynamic measurements were proposed by Kazemi et al. [12]–[14] and they showed that mangrove roots can reduce velocity by reducing fluid energy. They designed and evaluated experimental models and the flow structure. They demonstrated that physical models could reduce 90% of the maximum tidal current velocity. Kazemi et al. [12], [15], [16] also offered and the empirical equation for the drag coefficient to evaluate the aids of coastal tree root models in reducing coastal currents by tidal flows.

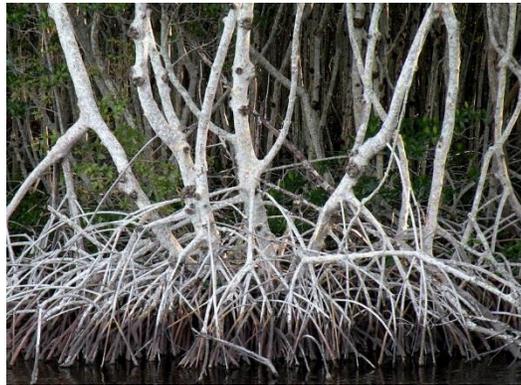

**Figure 1. Mangrove roots in Florida Everglades National Park.**

### III.  Methods

Most previous researches on the hydrodynamics of mangroves specify the complex passage generated in water over mangrove roots makes the flow friction dominated. Furukawa et al. calculated current and sediment transport in mangrove swamp at in  Australia and found that the mangrove roots produce complex two-dimensional flow fields and zones with root-scale turbulence. A high value of the drag coefficient was derived in the high dense mangrove vegetation. Mazda et al. used the momentum equation to find a balance between the water slope and drag force in and presented he drag coefficient of prop roots as a function of Reynolds number. The model proposed by Kazemi et al. [12–15], involves cylinders with a uniform diameter with a circular flow domain surrounds the cylinders. Water could penetrate the patch as a porous media at different velocities; The flow field around the patch was considered as two-dimensional flow since the heterogeneity of the flow field was negligible. They studied the flow in the wake of the patch, which consisted of arrays of circular cylinders. They observed a new concept that unlike for a solid obstruction, there is a delay in the inception of Von Kármán street due to the presence of an exit velocity from the trailing edge of the mangrove models. They confirmed it through the unsteady wake of the von Kármán vortex street by applying soap film flow visualization [19].

## IV.  Results and discussions

For the patch of circular cylinders, the physics of the flow is more complex compared to one single cylinder. The flow properties such as velocity inside and downstream of the patches for porosity change from φ=86%, to φ=96%, reduced around 40% of the tidal current velocity (20 cm/s) compared to a single standard cylinder. Contrary to the case of a standard cylinder, an unsteady wake region is formed toward the end of the patch model with to distinctive regions; high porosity followed by high turbulence and low porosity followed by low turbulence wake. Moreover, with an increase in the porosity, the flow interactions with the patch models turn into the weaker turblunce region. This eventually follows by a region of the steady wake length. Porosity generates steady wake length to grow longer considerably, and the vortex formation was impeded farther downstream. The patch porosity clearly affects the flow such that wakes are narrower and elongated. According to the patch models (Figure 2), patch drag coefficient decreases roughly 40% and 10% for φ=88% and φ=91%, respectively. This is contrary to a single cylinder, which is a good indication for the drag reduction with the addition of wake length because of porosity increase. This result has been verified by Chen et al. [13] and Kazemi et al.[19].

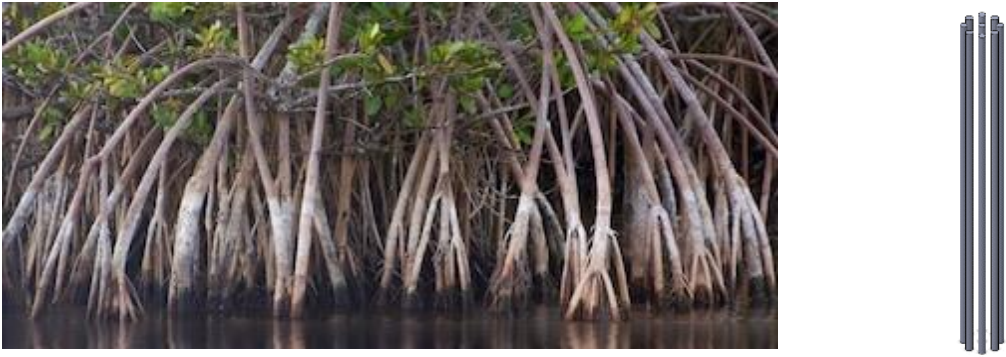

**Figure 2. Mangrove roots and a proposed patch model.**

Strouhal number describes the dominant non-dimensional wake shedding frequency. The Strouhal number surges with porosity but it is almost unbroken with Reynolds number, indicating a reliance of the vortex shedding frequency on the porosity. As the porosity escalates to 86%, the Strouhal number rises to 0.4 and marginally changes with Reynold number. The Strouhal number is roughly twice that of a standard single cylinder mainly due to the generated vortices as the made turbulence are followed along the core of vortical eddies, which help to recombine the vortices later downstream in the von Kármán vortex street. The results for different Reynolds numbers are compared with Kazemi et al [8] , furukawa et al. [17] , Zhang et al. [18] and Zong and nepf [19].

## V.  Conclusion

Human activities diminished the area of mangroves by 48% in the United States due to unexpected phenomena's such as the storms surges and tsunamis. Conserving or replanting coastal mangroves is a facile approach to be used to buffer communities from future tsunami events. Mangroves also enhance fisheries and forestry production that may be found in artificial coastal protection structures. Mangroves, however, are suitable for planting only on coastal mudflats and lagoons, which cover half of the continental coastline of the Atlantic Ocean. In a different place in the USA, the conservation of mangrove ecosystems or green belts of mangrove species can achieve the same protective role. This work reviewed the recent efforts on mangrove hydrodynamics. The geometrical parameters such as prop root height, porosity, root forms, length, diameter, and flexibility play a chief role. Our review suggests that mangroves attenuated tidal flow force and protected shorelines against damage.

## Acknowledgments

We would like to thank IHCC for the funding for this research and their useful feedback on our paper.